\RequirePackage{fix-cm}
\documentclass[twocolumn]{svjour3}          
\smartqed  
\usepackage{latexsym}
\usepackage{amsmath,amssymb}
\usepackage{bm}
\usepackage{graphicx}
\usepackage{hyperref}
\begin{document}

\title{Stochastic quantization approach to modeling of constrained Ito processes in discrete time
}


\author{Masakazu Sano}


\institute{M. Sano \at
              Current affiliation: \\
              Science Division, FreakOut, inc., \\ 
              Roppongi Hills Cross Point, 6-3-1 Roppongi, Minato-ku, Tokyo 106-0032, Japan \\
              \email{msano@fout.jp} 
}

\maketitle

\begin{abstract}
Stochastic quantization in physics has been considered to provide a path integral representation of a probability distribution for Ito processes. 
It has been indicated that the stochastic quantization can involve a potential term, if the Ito process is limited to Langevin equation. 
In this paper, in order to apply the stochastic quantization to engineering problems, we propose a novel method to incorporate a potential term into stochastic quantization of the general Ito process. 
This method indicates that weighted distribution gives rise to the potential term for the discrete-time path integral and 
 preserves the role of the path integral as the probability distribution, without making any assumptions on the drift term. 
A second order approximation on the stochastic fluctuations for the path integral gives difference equations which represent the time evolution of expectation value and covariance matrix for the stochastic processes. 
The difference equations explicitly derive Extended Kalman Filter and models on the constrained Ito processes by the identification of the potential function  with a penalty or barrier function.  
The numerical simulations of the constrained stochastic systems show that the potential term can constrain the nonlinear dynamics towards a minimum or a decreasing direction of the potential function.

\keywords{
	Stochastic quantization 
	\and Path integral
	\and Nonlinear dynamical systems 
	\and Extended Kalman Filter
	\and Penalty function 
	\and Barrier function}
\end{abstract}

\section{Introduction}

Nonlinear stochastic processes appear in a wide variety of research fields such as physics, biology, filtering, control, machine learning, finance.  
A deep understanding of such stochastic phenomena depends on the formulation of the stochastic processes and then various mathematical models has been considered.

In physics, euclidean path integral \cite{Kleinert:2004ev} has been well used as a powerful tool for modeling the stochastic processes and applied not only to physics but also to various areas. 
For example, the applications that had been discussed so far are as follows. 
Ref. \cite{RiskenFrank:1996} has explained that the path integral is able to satisfy a master equation of Langevin equation. 
Ref. \cite{Peliti:1985} has proposed a birth-death process represented by a path integral for an interacting system of stochastic fields on a lattice. 
The author of Ref. \cite{Kappen:2004,Kappen:2005} has discovered an unexpected relation between the path integral and stochastic nonlinear control.   
The relation has indicated that the Hamilton-Jacobi-Bellman (HJB) equation for Ito process involving a control input \cite{RFStengel:1994} 
becomes a linear partial differential equation (PDE), assuming a relation between the weight matrix of the control input and the covariance matrix of the Ito process. 
The solution of the PDE is given by the path integral which gives the control input. 
In \cite{Theodorou:2010-01,Theodorou:2010-02},  a task on finding a policy for a controlling robot in reinforcement learning has used the path integral method based on \cite{Kappen:2004,Kappen:2005}.
 As an application of the path integral to financial engineering \cite{Linetsky:1998,Baaquie:2004}, 
a pricing of a path-dependent option has considered the path integrals as Feynman-Kac formula satisfying a backward PDE. 
The path integral has involved a discount factor which deforms the stochastic dynamics of the option as the potential term.

For those stochastic models, the path integral is a solution of PDE on the stochastic process, 
however, in general, solving PDEs is quite difficult and solvable cases are limited. 
Therefore, it is natural to require a simple method which can directly define the path integral on a given stochastic process, in order to expand the applicabilities.  
The stochastic quantization (SQ) \cite{ParisiSourlas:1979,ParisiSourlas:1981,Parisi:1982ud,Parisi:1998,Damgaard:1987rr,Namiki:1992wf,Cooper:1994eh,Namiki:1993} 
has been known as such a direct method in physics. 
SQ has been originally a method to investigate an equilibrium state of Langevin dynamics in quantum physics. 
In \cite{Namiki:1992wf,Namiki:1993}, it has been pointed out that SQ gives a systematic and simple procedure 
defining the path integral as the probability distribution for a general Ito process, without solving the corresponding PDEs.
Using this fact, SQ has been applied a pricing of option in financial engineering \cite{Dash01:1988,Dash02:1989,DashBook:2004,Nakayama:2009}. 
If Ito process of a stock is given, SQ can define the path integral representation of the corresponding probability distribution which does not involve a potential term, directly. 
Such probability distribution can give the Feynman-Kac formula for the option which is a functional on a discount factor and a stock.

It is natural from this financial applications of SQ to consider whether SQ can define a path integral formula of probability distribution satisfying the dual forward PDE with a potential term or not.
The existence of such probability distribution gives the expectation that SQ can directly model the forward process affected from the external potential. 
If we identify the potential with a penalty or barrier function \cite{Fletcher:2000,Nocedal_Wright:2006}, 
we will be enable to discuss various models in which the potential function constrains the Ito process as the new applications of SQ on engineering.  
However, how to incorporate the potential term into SQ has been not clarified for the general Ito process. 
If the Ito process is restricted to Langevin equation which has the drift term consisting of a partial derivative of a scalar function, 
Lagrangian is able to involve a potential term by scaling of the path integral \cite{Parisi:1982ud,Parisi:1998}. 
The Langevin equation is quite restrictive for engineering problems, because Ito process appearing in engineering does not necessarily has such the drift term. 
Therefore, we require to clarify how the potential term can be incorporated in SQ without such the assumption on the drift term.

Taking into account this issue, in this paper, we will propose the introduction of weighted distribution \cite{Rao:1965_01,Rao:1984_01,Patil_Rao:1987_01} 
as a novel method to incorporate the potential term into the discrete-time SQ for the general Ito process. 
We will show the weighted distribution naturally induces the potential term in the path integral 
and preserves the role of the path integral as a probability distribution without imposing any assumptions on the drift term, 
although the weighted distribution has not been noticed in physics so far. 
For applications of SQ to engineering problems, we evaluate the path integral by a second order approximation on the stochastic fluctuations and then derives the difference equations.
Taking an appropriate potential function, the difference equations will be applied to the modeling of filtering and control and to the numerical simulations of the control models.

This paper is organized as follows. 
In section \ref{Stochastic quantization for nonlinear control}, 
we will introduce SQ using weighted distribution in discrete time.  
The existence of the dual backward formula is mentioned. 
In section \ref{The state-space representation}, 
the path integral gives the difference equations as the evolution equations of the expected value and the covariance matrix on the stochastic process, applying a second order approximation on the stochastic fluctuations.
In section \ref{Applications to the Extended Kalman Filter and nonlinear stochastic  control models}, 
by taking the potential function as the penalty or barrier function, the difference equations give rise to Extended Kalman Filter (EKF) \cite{Simon:2006,Saerkkae:2013} and models on nonlinear stochastic control. 
We also discuss numerical simulations of the nonlinear stochastic control. 
The section \ref{conclusion} will be devoted to the conclusions.

\section{SQ and Weighted distribution}\label{Stochastic quantization for nonlinear control}

We will introduce SQ to define a probability distribution on the Ito process by path integral. 
A  weight function of the weighted distribution gives an additional potential term in the path integral and 
preserves the role of the path integral as the probability distribution, without imposing any assumptions on the drift term. 
To clarify the difference with the typical SQ discussed in physics, we will investigate PDE that the probability distribution involving the weighted distribution satisfies in continuous time. 
It is shown that the PDE recovers the Fokker-Planck equation, if the potential term vanishes. 
Although the Fokker-Planck equation is able to induce an additional potential term by a scaling of the solution and 
by assuming that the drift term of the Ito process is proportional to a partial derivative of a scalar function, 
we find that the SQ with the weighted distribution can introduce the additional potential term without the scaling and the assumption.

\subsection{Path integral formula of SQ}

We will consider a finite-time system in discrete time.  $\bm{\eta}_{t}$ is a $m$-dimensional random variable which obeys the normal (Gaussian) distribution with 
$E[ \bm{\eta}_{t} ] = 0$ and  $E[ \bm{\eta}_{t} \bm{\eta}_{t}^{T} ] = \bm{1}_{m \times m}$. 
The probability distribution is given by 
\begin{align}
	1 &= \int^{+ \infty}_{- \infty} d^{m} \eta_{t} \Bigl( \frac{\varDelta t}{2 \pi } \Bigr)^{m/2}
		 \exp{ \Bigl( - \frac{1}{2} \bm{\eta}^{T}_{t} \bm{\eta}_{t} \varDelta t \Bigr) }. \label{stnc_01} 
\end{align}
Using $\bm{\eta}_{t}$,  a discrete-time Ito process is defined by 
\begin{align}
	\frac{\bm{x}_{t + \varDelta t} - \bm{x}_t}{\varDelta t} -\bm{f}_{t}( \bm{x}_t ) = \bm{E}^{-1}_{t} \bm{\eta}_{t} \label{stnc_02}
\end{align}
where   $\bm{x}_t$ , $\bm{f}_{t}( \bm{x}_t )$ are $m$-dimensional vectors.  
$\bm{E}^{-1}_{t}$ is  an $m \times m$-matrix which has the inverse matrix $\bm{E}_{t}$ with $\bm{E}^{-1}_{t}\bm{E}_{t} = \bm{1}_{m \times m}$.  
The Ito process (\ref{stnc_02}) indicates that $\bm{x}_{t + \varDelta t}$ is a random variable and $\bm{x}_{t}$ is already given at $t$.

SQ \cite{ParisiSourlas:1979,ParisiSourlas:1981,Parisi:1982ud,Parisi:1998,Damgaard:1987rr,Namiki:1992wf,Cooper:1994eh,Namiki:1993} 
has been known as a technique which defines the probability distribution of the Ito process (\ref{stnc_02}) by path integral.  
To define the probability distribution for Eq. (\ref{stnc_02}),  the SQ introduces the following delta function:   
\begin{align}
	1 &= \int^{+ \infty}_{- \infty} d^{m} x_{t+\varDelta t} \frac{\sqrt{g_{t}} }{(\varDelta t)^{m}} \notag \\[10pt]
		&\qquad \qquad \times \delta \Biggl( \bm{E}_{t} \Bigl( \frac{ \bm{x}_{t + \varDelta t} -\bm{x}_t}{\varDelta t} -\bm{f}_{t}( \bm{x}_t) \Bigr) -  \bm{\eta}_{t} \Biggr) \label{eq:02-02-03}
\end{align}
where $\bm{g}_{t} \equiv \bm{E}^{T}_{t} \bm{E}_{t} > 0$ and $\sqrt{g_{t}} \equiv \det \bm{E}_{t}$. 
We assume that an initial probability density of $\bm{x}_{t}$ is given by a positive function $P(\bm{x}_{t}, t) \geq 0$ satisfying the following equation; 
\begin{align}
	1 &=  \int^{+ \infty}_{- \infty} d^{m}x_{t} P(\bm{x}_{t}, t). \label{eq:02-05}
\end{align}
Inserting  Eq. (\ref{stnc_01}) and E.  (\ref{eq:02-02-03}) to Eq.  (\ref{eq:02-05}), we obtain the following equation after the integration on $\bm{\eta}_{t}$: 
\begin{align}
	1 = \int^{+ \infty}_{- \infty} d^{m} x_{t + \varDelta t} p(\bm{x}_{t + \varDelta t}, t + \varDelta t)  \label{eq:02-8}
\end{align}
where 
\begin{align}
	 &p(\bm{x}_{t + \varDelta t}, t + \varDelta t) \notag \\[10pt]
	 &\qquad \equiv \int^{+ \infty}_{- \infty} d^{m} x_{t}  
	 	\frac{\sqrt{g_{t}}}{(2 \pi \varDelta t)^{m/2}}  e^{-  T_{t  + \varDelta t }  \varDelta t}
			P(\bm{x}_{t}, t),  \label{eq:02-07} \\[10pt]
	&T_{t + \varDelta t} = \frac{1}{2}  
		\Bigl( \frac{\bm{x}_{t + \varDelta t} - \bm{x}_t}{\varDelta t} -\bm{f}_{t}( \bm{x}_t) \Bigr)^{T} \notag \\[10pt]
				&\qquad \qquad \qquad \times \bm{g}_{t} 
				\Bigl( \frac{\bm{x}_{t + \varDelta t} - \bm{x}_t}{\varDelta t} -\bm{f}_{t}( \bm{x}_t) \Bigr).  \label{eq:02-04}
\end{align}
$T_{t + \varDelta t}$ is the kinetic term. 
Eq. (\ref{eq:02-07}) is the path integral representation of the probability distribution for the Ito process (\ref{stnc_02}) 
and describes an evolution of the initial distribution $P(\bm{x}_{t}, t)$. 
If we impose 
\begin{align}
	p(\bm{x}_{t + \varDelta t}, t + \varDelta t)  = P(\bm{x}_{t + \varDelta t}, t + \varDelta t), \label{assumption_fo_prob_01}
\end{align} 
Eq. (\ref{eq:02-07}) is the path integral solution for the Fokker-Planck equation \cite{RiskenFrank:1996}. 
In order to constrain the stochastic dynamics of the Ito process (\ref{stnc_02}) for the applications to various engineering problems, 
we have to introduce external forces without  the condition (\ref{assumption_fo_prob_01}) and any assumption for the drift term of Eq. (\ref{stnc_02}). 
By facts on the path integral \cite{Kleinert:2004ev}, it is expected that introducing of an additional potential term in the path integral induces such forces. 
In the following discussions, we introduce weighted distribution to naturally define the potential term in the path integral.

\subsection{The introduction of  weighted distribution}

We will consider an extension of Eq. (\ref{eq:02-07}) to path integral with an additional potential function without assumption for the drift term of the Ito process (\ref{stnc_02}). 
Inspired by the weighted distribution \cite{Rao:1965_01,Rao:1984_01,Patil_Rao:1987_01} which has been considered in the area of statistics,  
we would like to introduce a weight function, 
\begin{align}
	\exp( - U_{ t+\varDelta t}(\bm{x}_{t+\varDelta t}) \varDelta t ). 
\end{align} 
Using this weight function, we define $P(\bm{x}_{t + \varDelta t}, t + \varDelta t)$ as a weighted distribution: 
\begin{align}
	 &P(\bm{x}_{t + \varDelta t}, t + \varDelta t) \notag \\[10pt]
	 &\equiv e^ {- U_{ t+\varDelta t}(\bm{x}_{t+\varDelta t}) \varDelta t  } p(\bm{x}_{t + \varDelta t}, t + \varDelta t)  \label{eq:02-11}  \\[10pt]
	 &= \int^{+ \infty}_{- \infty} d^{m} x_{t} 
	 \frac{\sqrt{g_{t}}}{(2 \pi \varDelta t)^{m/2}} \notag \\[10pt]
	 &\qquad \qquad \times e^{- ( T_{t + \varDelta t} + U_{ t+\varDelta t}(\bm{x}_{t+\varDelta t}) ) \varDelta t}
	 	 P(\bm{x}_{t}, t). \label{eq:02-12} 
\end{align}
We impose conservation of probability for the weighted distribution as follows: 
\begin{align}
	1 =  \int^{+ \infty}_{- \infty} d^{m} x_{t + \varDelta t} P( \bm{x}_{t + \varDelta t}, t + \varDelta t). \label{pc}
\end{align}
This condition implies that the transition probability distribution for Eq. (\ref{eq:02-12}) can be defined by 
\begin{align}
	&P( \bm{x}_{t + \varDelta t}, t + \varDelta t | \bm{x}_{t}, t) \notag \\[10pt]
	&\qquad =  \frac{\sqrt{g_{t}}}{(2 \pi \varDelta t)^{m/2}} 
		 \times e^{- ( T_{t + \varDelta t} + U_{ t+\varDelta t}(\bm{x}_{t+\varDelta t}) ) \varDelta t} \label{trans}
\end{align}
where we have assumed that the normalization is involved in the potential function. 
Therefor, Eq. (\ref{eq:02-12}) is interpreted as the evolution of the initial probability density Eq. (\ref{eq:02-05}) through the transition probability distribution Eq. (\ref{trans}). 

As shown in \cite{ParisiSourlas:1981,Parisi:1982ud,Parisi:1998}, SQ without the weighted distribution can induce a potential term 
by introducing an additional drift term, after a scaling of the probability density which satisfies Fokker-Planck equation. 
To understand that our model (\ref{eq:02-12}) is different with the case in \cite{ParisiSourlas:1981,Parisi:1982ud,Parisi:1998}, 
we will consider the continuous case of Eq. (\ref{eq:02-12}). 
Considering a fluctuation $\bm{\xi}_{t} = \bm{x}_{t} - \bm{x}_{t + \varDelta t}$ and 
Expanding Eq.  (\ref{eq:02-12}) around $\bm{x}_{t + \varDelta t}$ by $\bm{x}_{t} = \bm{x}_{t + \varDelta t} + \bm{\xi}_{t}$, 
the continuous-time limit $\varDelta t \rightarrow 0$, after the integral on $\bm{\xi}_{t}$, gives the following equation: 
\begin{align}
	\partial_{t} P(\bm{x}_{t}, t) 
		=&\sum_{\mu, \nu = 1 }^{m} \frac{1}{2} g_{t}^{\mu \nu} \partial_{\mu} \partial_{\nu} P(\bm{x}_{t}, t) \notag \\[10pt]
			&- \sum_{\mu = 1}^{m} \partial_{\mu} ( f^{\mu}_{t} (\bm{x}_{t} ) P(\bm{x}_{t}, t) ) 
			- U_{t}( \bm{x}_{t} ) P(\bm{x}_{t}, t) \label{eq:efp_01} 
\end{align}
where $g_{t}^{\mu \nu} \equiv (\bm{g}_{t}^{-1})^{\mu\nu}$. 
If $U_{t}( \bm{x}_{t} ) = 0$, Eq. (\ref{eq:efp_01}) is exactly equivalent to the Fokker-Planck equation. 
Therefore, Eq. (\ref{eq:02-12}) is a simple extension of the typical SQ. 
For a case of $\bm{f}_{t} = \bm{0}$ in Eq. (\ref{eq:efp_01}), we also find that $U_{t}( \bm{x}_{t} ) $ has the same role as potential function in physics, 
because  Eq. (\ref{eq:efp_01}) becomes the euclidean Schroedinger equation. 
In order to understand the difference, we will consider the following scaling of the probability distribution: 
\begin{align}
	P(\bm{x}_{t}, t) = e^{W(\bm{x}_{t})} \rho (\bm{x}_{t}, t) \label{eq:efp_02} 
\end{align}
where $W(\bm{x}_{t})$ is a scalar function. 
By the scaling, Eq. (\ref{eq:efp_01}) becomes 
\begin{align}
	&\partial_{t} \rho (\bm{x}_{t}, t) \notag \\[10pt]
		&= \sum_{\mu, \nu = 1 }^{m}  \frac{1}{2} g_{t}^{\mu \nu} \partial_{\mu} \partial_{\nu} \rho (\bm{x}_{t}, t) \notag \\[10pt]
			&\quad - \sum_{\mu = 1 }^{m}  \partial_{\mu} \{ (f^{\mu}_{t} (\bm{x}_{t} ) -   \sum_{\nu = 1 }^{m} g_{t}^{\mu\nu}\partial_{\nu} W (\bm{x}_{t}) ) \rho (\bm{x}_{t}, t) \} \notag \\[10pt]
			&\quad + \sum_{\mu = 1 }^{m} (f^{\mu}_{t} (\bm{x}_{t} ) - \sum_{\nu = 1 }^{m}  g_{t}^{\mu\nu} \partial_{\nu} W (\bm{x}_{t}) ) \partial_{\mu} W (\bm{x}_{t}) \rho (\bm{x}_{t}, t)  \notag \\[10pt]
			&\quad - \Bigl [ U_{t}( \bm{x}_{t} )  
			+ \frac{1}{2} \sum_{\mu, \nu = 1 }^{m} \{ g_{t}^{\mu \nu} \partial_{\mu} \partial_{\nu}W (\bm{x}_{t}) \notag \\[10pt]
				&\qquad \qquad \qquad + g_{t}^{\mu \nu} \partial_{\mu} W (\bm{x}_{t}) \partial_{\nu} W (\bm{x}_{t}) \} \Bigr ] \rho (\bm{x}_{t}, t). \label{eq:efp_03}
\end{align}
If we take additional condition on $\bm{f}_{t}$ as follows: 
\begin{align}
	f^{\mu}_{t} (\bm{x}_{t} ) - \sum_{\nu = 1 }^{m} g_{t}^{\mu\nu} \partial_{\nu} W (\bm{x}_{t}) = 0,  \label{eq:efp_04}
\end{align}
Eq. (\ref{eq:efp_03}) is reduced to the following equation \cite{ParisiSourlas:1981,Parisi:1982ud,Parisi:1998}: 
\begin{align}
	&\partial_{t} \rho (\bm{x}_{t}, t) \notag \\[10pt]
	&= \sum_{\mu, \nu = 1 }^{m}  \frac{1}{2} g_{t}^{\mu \nu} \partial_{\mu} \partial_{\nu} \rho (\bm{x}_{t}, t) -   U_{t}( \bm{x}_{t} )  \rho (\bm{x}_{t}, t)   \notag \\[10pt]
	&\quad - \frac{1}{2} \sum_{\mu, \nu = 1 }^{m}
		( g_{t}^{\mu \nu} \partial_{\mu} \partial_{\nu} W + g_{t}^{\mu \nu} \partial_{\mu} W  \partial_{\nu} W )  \rho (\bm{x}_{t}, t).   \label{parisi_diff}
\end{align}
We find that Eq. (\ref{parisi_diff}) is formally equivalent to the euclidean Schroedinger equation. 
For $U_{t}( \bm{x}_{t} )=0$, this fact indicates that the Fokker-Plank equation can introduce the potential function under the assumptions (\ref{eq:efp_04}). 
However, this potential function and the condition (\ref{eq:efp_04}) are quite restrictive for applications on various engineering probrems, 
because a realistic stochastic process does not necessarily have such the scalar function $W(\bm{x}_{t})$ satisfying Eq. (\ref{eq:efp_04}). 
In our model, we can consider the path integral involving the potential function $U_{t}( \bm{x}_{t} )$ 
for a general stochastic process, without the assumption (\ref{eq:efp_04}). 
Therefore, if an appropriate potential function $U_{t}( \bm{x}_{t} ) $ is taken for an engineering problem, it is expected 
that we can constrain a general stochastic process toward a desired path for solving a task on the problem. 
This point is the major difference from SQ in physics.

\subsection{The existence of dual backward formula}

Finally, we will mention whether a dual backward formula can be defined from the forward formula (\ref{eq:02-12}).  
If the dual function at $t_{N} \equiv t + N \varDelta t$ is given by  a real function $C_{ t_{N} }(\bm{x}_{t_{N}})$ which is not necessarily normalized, 
an inner product for (\ref{eq:02-12}) and $C_{ t }(\bm{x}_{t}) $ at $t_{N}$ can be represented by 
\begin{align}
	&\int^{+ \infty}_{- \infty} d^{m} x_{t_{N}} C_{ t_{N} }(\bm{x}_{t_{N}}) P( \bm{x}_{t_{N}}, t_{N}) \notag \\[10pt]
	&\quad =\int^{+ \infty}_{- \infty} d^{m} x_{t} C_{ t }(\bm{x}_{t}) P( \bm{x}_{t}, t) \label{d_01}
\end{align}
where $C_{ t }(\bm{x}_{t })$ has been defined as follows: 
\begin{align}
	&C_{ t }(\bm{x}_{t }) \notag \\[10pt]
	&\equiv \int^{+ \infty}_{- \infty} d^{m} x_{t + \varDelta t} C_{ t + \varDelta t }(\bm{x}_{ t + \varDelta t }) P( \bm{x}_{ t + \varDelta t }, t + \varDelta t | \bm{x}_{t}, t) \label{d_02} \\[10pt]
	&= \int^{+ \infty}_{- \infty} d^{m} x_{t_{N}} C_{t_{N}}(\bm{x}_{t_{N}})  \notag \\[10pt]
	&\qquad\times \prod_{i=t}^{t_{N} - \varDelta t} d^{m} x_{i} P( \bm{x}_{i + \varDelta t}, i +\varDelta t | \bm{x}_{i}, i). \label{d_03}
\end{align}
The function $C_{ t }(\bm{x}_{t }) $ clearly satisfies the backward Kolmogolov equation (\ref{d_02}). 
If $C_{ t_{N} }(\bm{x}_{ t_{N} })$ is taken as a functional on options, time and an appropriate normalization constant, 
Eq. (\ref{d_03}) is able to be interpreted as the Feynman-Kac formula involving a discount factor \cite{Linetsky:1998,Baaquie:2004} 
proportional to $- \sum_{i = t}^{t_{N} - \varDelta t} U_{i +  \varDelta t}(\bm{x}_{i +  \varDelta t}) \varDelta t$ by (\ref{trans}).

We would like to also mention a difference with a backward path integral of a nonlinear control model discussed in \cite{Kappen:2004,Kappen:2005}.
If we assume that $C_{ t_{N} }(\bm{x}_{ t_{N} })$ is an exponential function on a cost function like $\exp(-\phi(\bm{x}_{t_{N}}) \varDelta t)$, 
$C_{ t }(\bm{x}_{t })$ gives a similar path integral formula on a cost-to-go function in \cite{Kappen:2004,Kappen:2005}. 
However, in discrete time, Ref. \cite{Kappen:2004,Kappen:2005} uses the potential function at $t$ instead of the potential function at $t + \varDelta t$. 
Therefore, in the discrete-time level, the model proposed in this paper is directly different because a different discretization for the potential function in term of the time variable is applied in \cite{Kappen:2004,Kappen:2005}.

\section{Nonlinear Difference equations}\label{The state-space representation}

We have discussed SQ with the weight distribution in the previous section. 
It has been indicated without assumption like Eq. (\ref{eq:efp_04}) that the weight function induces the potential function in the path integral, 
preserving the role as the probability distribution. 
By analogy with the path integral in physics, we have expected a constrained dynamics by a forces derived from the potential function. 
To confirm such expectations through explicit examples involving the numerical simulations, 
we would like to evaluate the path integral by a second order expansion on the stochastic fluctuations.
We will derive the difference equations as evolution equations of covariance matrix and expectation value for the stochastic variables.

As the initial probability distribution (\ref{eq:02-05}), we will take gaussian distribution with the mean value $\bm{\hat{x}}_{t | t}$ and the covariance matrix $\hat{\varSigma}_{t | t }$:  
\begin{align}
	P(\bm{x}_{t}, t  ) 
		&= \frac{1}{(2 \pi )^{m/2} | \hat{\varSigma}_{t | t}| ^{1/2} }
			e^ {- \frac{1}{2} \delta \bm{x}_{t}^{T} 
			\hat{\varSigma}_{t | t}^{-1}  \delta \bm{x}_{t} }  \label{eq:05-01}
\end{align}
where the stochastic fluctuation has been defined by
\begin{align}
	\delta \bm{x}_{t } &=  \bm{x}_{t } - \bm{\hat{x}}_{t | t }. 
\end{align}
To perform the integration in Eq. (\ref{eq:02-07}),  we will expand $\bm{x}_{t} + \bm{f}_{t}(\bm{x}_{t}) \varDelta t$ around $\bm{\hat{x}}_{t | t}$ as follows: 
\begin{align}
	\bm{x}_{t} + \bm{f}_{t}(\bm{x}_{t}) \varDelta t \simeq 
	\bm{\hat{x}}_{t  | t } + \bm{f}_{t}(\bm{\hat{x}}_{t| t }) \varDelta t + \bm{F}_{t} \delta \bm{x}_{t} \label{eq:05-18-00}
\end{align}
where we have defined 
\begin{align}
	(\bm{F}_{t})_{~\nu}^{\mu} \equiv \delta^{\mu}_{~\nu} +  \partial_{\nu} f^{\mu} (\bm{\hat{x}}_{t | t } ) \varDelta t.
\end{align}
After the integration on $\delta \bm{x}_{t}$ ($d \delta \bm{x}_{t} = d \bm{x}_{t}$), Eq. (\ref{eq:02-07})  is represented by 
\begin{align}
	&p(\bm{x}_{t+  \varDelta t }, t + \varDelta t ) \notag \\[10pt]
		&=  \frac{1}{(2 \pi )^{m/2} | \hat{\varSigma}_{t  + \varDelta t | t } |^{1/2} }
			 \times  e^{ - \frac{1}{2} \delta \bm{x}_{t + \varDelta t | t} ^{T} 
				\hat{\varSigma}_{t + \varDelta t | t}^{-1}  \delta \bm{x}_{t + \varDelta t | t} } \label{eq:05-18}
\end{align}
where $\delta \bm{x}_{t + \varDelta t | t} =  \bm{x}_{t + \varDelta t } - \bm{\hat{x}}_{t +\varDelta t| t }$ and the prediction processes have been defined by 
\begin{align}
	\bm{\hat{x}}_{t+\varDelta t | t}  &
		= \bm{\hat{x}}_{t | t } +  \bm{f}_{t} ( \bm{\hat{x}}_{t | t }  ) \varDelta t, \label{eq:03-02} \\[10pt]
	 \hat{\varSigma}_{t + \varDelta t | t}  & 
	 	= \bm{F}_{t} \hat{\varSigma}_{t | t }   \bm{F}_{t}^{T} 
			+ \bm{g}^{-1}_{t} \varDelta t.   \label{eq:03-03}
\end{align}

In order to evaluate the path integral (\ref{eq:02-12}), 
we are going to assume that the potential function $U_{t + \varDelta t}$ is defined as follows: 
\begin{align}
	U_{t + \varDelta t}( \bm{x}_{t+\varDelta t} )  &= 
		V^{eff}_{t + \varDelta t}( \bm{x}_{t+\varDelta t} ) 
		- \frac{1}{\varDelta t}  \ln N_{t + \varDelta t},
		 \label{eq:05-02} \\[10pt]
	V^{eff}_{t + \varDelta t} ( \bm{x}_{t+\varDelta t} )  &= V_{t + \varDelta t} ( \bm{l}_{t + \varDelta t} (\bm{x}_{t + \varDelta t}) ) \notag \\[10pt]
		&\qquad + \delta_{c} V_{t + \varDelta t} ( \bm{x}_{t + \varDelta t} ) \label{eq:05-02-02} 
\end{align}
where  $V^{eff}_{t + \varDelta t}$ is the effective potential, 
$\delta_{c} V_{t + \varDelta t}( \bm{x}_{t + \varDelta t} )$ is a counter term which is decided in later discussion and $N_{t + \varDelta t}$ is a normalization. 
In quantum mechanics, it is known that path integrals are able to take in the quantum fluctuations around the classical paths by performing the integral on the fluctuations \cite{Kleinert:2004ev}.
To apply this method for evaluating Eq. (\ref{eq:02-12}), 
we will consider $V_{t + \varDelta t}$ up to the second order terms on $\delta \bm{x}_{t + \varDelta t | t}$ as follows: 
\begin{align}
	&V_{t + \varDelta t} ( \bm{l}_{t+1} ( \bm{x}_{t + \varDelta t}) ) \notag \\[10pt]
		&\quad \simeq \Bigl\{  
			V_{t + \varDelta t}(\bm{l}_{t + \varDelta t}) \notag \\[10pt]
		& \qquad \qquad  - \delta \bm{x}_{t + \varDelta t | t}^{T} \bm{H}_{t + \varDelta t}^{T} \bm{\nabla}_{\bm{l}_{t + \varDelta t}} V_{t + \varDelta t}  \notag \\[10pt]
		& \qquad \qquad + \frac{1}{2} \delta \bm{x}_{t + \varDelta t | t}^{T} \bm{H}_{t + \varDelta t}^{T} \varSigma_{\nu_{t + \varDelta t}}^{-1} \bm{H}_{t + \varDelta t}
		  	  \delta \bm{x}_{t + \varDelta t | t} 
		\Bigr\}  \notag \\[10pt]
			& \qquad \qquad+ \frac{1}{2} \delta \bm{x}_{t + \varDelta t | t}^{T}  {\varSigma'}_{\nu_{t + \varDelta t}}^{-1} 
		  	  \delta \bm{x}_{t + \varDelta t | t} \label{eq:05-04}
\end{align}
where we have defined the following equations:
\begin{align}
	&\bm{l}_{t + \varDelta t} 
		\equiv \bm{l}_{t + \varDelta t}(\bm{\hat{x}}_{t + \varDelta t | t }) , \label{eq:05-08} \\[10pt]
	&\bm{\nabla}_{\bm{l}_{t + \varDelta t}} V_{t + \varDelta t} 
	 	\equiv  \bm{\nabla}_{\bm{l}_{t + \varDelta t}} V_{t + \varDelta t} |_{\bm{x}_{t + \varDelta t} = \bm{\hat{x}}_{t + \varDelta t | t }}, \label{eq:05-10} \\[10pt]
	&(\bm{H}_{t + \varDelta t}^{T})_{~~\mu}^{m}
		 \equiv - \partial_{\mu} l^{m}_{t + \varDelta t} |_{\bm{x}_{t + \varDelta t} = \bm{\hat{x}}_{t + \varDelta t | t }} \notag \\[10pt]
	 	&\quad\quad\quad\quad\quad  \equiv - (\bm{\nabla} \bm{l}_{t + \varDelta t})_{\mu}^{~m},  \label{eq:05-11} \\[10pt]
	&(\varSigma_{\nu_{t + \varDelta t}}^{-1})_{mn}
		 \equiv \partial_{l^{m}_{t + \varDelta t}} \partial_{l^{n}_{t + \varDelta t}}  V_{t + \varDelta t} |_{\bm{x}_{t + \varDelta t} = \bm{\hat{x}}_{t + \varDelta t | t }}, \label{eq:05-14} \\[10pt]
	&({\varSigma'}_{\nu_{t + \varDelta t}}^{-1})_{\mu\nu}
		 \equiv \sum_{m=1}^{m}  \partial_{\mu} \partial_{\nu} l^{m}_{t + \varDelta t} \notag \\[10pt]
		&\qquad\qquad\qquad\qquad \times \partial_{l^{m}_{t + \varDelta t}}  
		 	V_{t + \varDelta t}  |_{\bm{x}_{t + \varDelta t} = \bm{\hat{x}}_{t  + \varDelta t| t }}. \label{eq:05-15}
\end{align}
We also take the normalizations $N_{t + \varDelta t}$ as follows:   
\begin{align}
	N_{t + \varDelta t} 
	\equiv & | \varSigma_{\nu_{t + \varDelta t}} / \varDelta t +  \bm{H}_{t + \varDelta t} \hat{\varSigma}_{t + \varDelta t| t }   \bm{H}_{t + \varDelta t}^{T} |^{1/2} \notag \\[10pt]
	&\times |\varSigma_{\nu_{t + \varDelta t}}|^{ -1/2}  (\varDelta t)^{m/2}  \times \exp  ( \mathcal{N} _{t + \varDelta t} \varDelta t  ) \label{eq:05-16} 
\end{align}
where $\mathcal{N}_{t + \varDelta t}$ is defined by 
\begin{align}
	\mathcal{N}_{t + \varDelta t} 
	\equiv &  V_{t + \varDelta t}(\bm{l}_{t + \varDelta t}) \notag \\[10pt]
		&- \frac{1}{2} ( \bm{\nabla}_{\bm{l}_{t + \varDelta t}} V_{t + \varDelta t} )^{T} \bm{H}_{t + \varDelta t} \notag \\[10pt]
			&\times  ( \hat{\varSigma}_{t + \varDelta t | t }^{-1} + \bm{H}_{t + \varDelta t}^{T}  \varSigma_{\nu_{t + \varDelta t}}^{-1} \bm{H}_{t + \varDelta t} \varDelta t  )^{-1} \notag \\[10pt]
				 	&\times \bm{H}_{t + \varDelta t}^{T}  \bm{\nabla}_{\bm{l}_{t + \varDelta t}}  V_{t + \varDelta t} \varDelta t. \label{eq:05-17}
\end{align}
For a stability of the numerical simulations, we will assume
\begin{align}
	& \varSigma_{\nu_{t + \varDelta t}}^{-1} > 0, \label{eq:05-17-01} \\[10pt]
	& \delta_{c} V_{t + \varDelta t} ( \bm{x}_{t + \varDelta t} ) \equiv - \frac{1}{2} (\delta \bm{x}_{t + \varDelta t | t})^{T}  {\varSigma'}_{\nu_{t + \varDelta t}}^{-1} 
		  	  \delta \bm{x}_{t + \varDelta t | t}. \label{eq:05-17-02}
\end{align}
Using  Eq. (\ref{eq:05-16}), Eq. (\ref{eq:05-17}), Eq. (\ref{eq:05-17-01}), Eq.  (\ref{eq:05-17-02})  and Eq. (\ref{eq:A02-01}), it is found that the weighted distribution is given by
\begin{align}
	&P( \bm{x}_{t + \varDelta t}, t + \varDelta t  ) \notag \\[10pt]
	&= \frac{1}{(2 \pi )^{m/2} | \hat{\varSigma}_{t + \varDelta t | t + \varDelta t}| ^{1/2} }
			e^ {- \frac{1}{2} \delta \bm{x}_{t + \varDelta t}^{T} 
			\hat{\varSigma}_{t + \varDelta t | t + \varDelta t}^{-1}  \delta \bm{x}_{t + \varDelta t } } 
	\label{eq:05-27-01} 
\end{align}
where $ \bm{\hat{x}}_{t + \varDelta t | t+ \varDelta t  } $ and $ \hat{\varSigma}_{t  + \varDelta t | t + \varDelta t }$ are defined as follows: 
\begin{align}
	\bm{\hat{x}}_{t + \varDelta t | t  + \varDelta t} 
		&= \int_{-\infty}^{\infty} d^{m}x_{t + \varDelta t}  P( \bm{x}_{t + \varDelta t}, t + \varDelta t  ) \bm{x}_{t + \varDelta t}  \notag \\[10pt]
		&= \bm{\hat{x}}_{t + \varDelta t | t} \notag \\[10pt]
			&\quad+ \hat{\varSigma}_{t + \varDelta t | t }  \bm{H}_{t + \varDelta t}^{T} \notag \\[10pt]
				 &\quad\times (\varSigma_{\nu_{t + \varDelta t}}^{ } / \varDelta t  + \bm{H}_{t + \varDelta t} \hat{\varSigma}_{t + \varDelta t | t } \bm{H}_{t + \varDelta t}^{T} )^{-1} \notag \\[10pt]
				 &\quad \times  \varSigma_{\nu_{t + \varDelta t}} \bm{\nabla}_{\bm{l}_{t + \varDelta t}} V_{t + \varDelta t},
			 \label{eq:03-04} \\[10pt]
	\hat{\varSigma}_{t + \varDelta t | t  + \varDelta t} 
		&= \int_{-\infty}^{\infty} d^{m}x_{t + \varDelta t} P( \bm{x}_{t + \varDelta t}, t + \varDelta t  ) \notag \\[10pt]
			 &\quad \times (\bm{x}_{t + \varDelta t} - \bm{\hat{x}}_{t + \varDelta t | t  + \varDelta t} ) \notag \\[10pt]
			&\quad  \times ( \bm{x}_{t + \varDelta t} - \bm{\hat{x}}_{t + \varDelta t | t  + \varDelta t} )^{T}   \notag \\[10pt]
		&= \hat{\varSigma}_{t + \varDelta t | t} \notag \\[10pt]
			&\quad - \hat{\varSigma}_{t + \varDelta t | t } \bm{H}_{t + \varDelta t}^{T} \notag \\[10pt]
			 &\quad \times ( \varSigma_{\nu_{t + \varDelta t}}^{ }  / \varDelta t
				+ \bm{H}_{t + \varDelta t} \hat{\varSigma}_{t + \varDelta t | t}   \bm{H}_{t + \varDelta t}^{T}  )^{-1} \notag \\[10pt]
				& \quad \times \bm{H}_{t + \varDelta t} \hat{\varSigma}_{t + \varDelta t | t}. \label{eq:03-05}
\end{align}
Eq. (\ref{eq:03-02}), Eq. (\ref{eq:03-03}), Eq. (\ref{eq:03-04}) and Eq. (\ref{eq:03-05}) are difference equations that we would like to derive. 

Finally, we will comment on the assumptions Eq. (\ref{eq:05-17-01}) and Eq. (\ref{eq:05-17-02}). 
If we does not assume Eq. (\ref{eq:05-17-01}) and Eq. (\ref{eq:05-17-02}), the inverse matrix in Eq. (\ref{eq:A02-02-02}) and Eq. (\ref{eq:A02-03-02}) is replaced by
\begin{align}
	( \hat{\varSigma}_{t  + \varDelta t | t }^{-1} + \bm{H}_{t + \varDelta t }^{T} \varSigma_{\nu_{t + \varDelta t }}^{ -1}  \bm{H}_{t + \varDelta t } \varDelta t  
		+  {\varSigma'}_{\nu_{t + \varDelta t }}^{ -1}  \varDelta t )^{-1}. \notag 
\end{align} 
We are able to ignore the second and the third term of the inverse matrix in the continuous limit $\varDelta t \rightarrow 0$.
However, for the discrete-time, this inverse matrix has a possibility of the divergence or of the negative value because of a case where 
$\varSigma_{\nu_{t + \varDelta t }}^{ -1}$ and ${\varSigma'}_{\nu_{t + \varDelta t }}^{ -1}$ are
singular or negative, in general. 
This causes a difficulty of the stable numerical simulations. 
Therefore, we will impose Eq. (\ref{eq:05-17-01}) and Eq. (\ref{eq:05-17-02}) through this paper.

\section{Applications to filtering and control}\label{Applications to the Extended Kalman Filter and nonlinear stochastic  control models}

In the previous section, we have derived the difference equations on the evolution of the nonlinear stochastic systems involving the potential function. 
The difference equation Eq. (\ref{eq:03-04}) implies that the expectation value $ \bm{\hat{x}}_{t  + \varDelta t | t } $ is affected by forces from the potential at $t + \varDelta t$. 
To understand this, more explicitly, note that 
by Eq. (\ref{eq:A02-02-02})  in Appendix \ref{A02}, Eq. (\ref{eq:03-04})  can be represented as follows: 
\begin{align}
	\bm{\hat{x}}_{t + \varDelta t | t  + \varDelta t } 
		&= \bm{\hat{x}}_{t  + \varDelta t | t } \notag \\[10pt]
			&\quad - ( \hat{\varSigma}_{t  + \varDelta t | t}^{-1} + \bm{H}_{t + \varDelta t}^{T} \varSigma_{\nu_{t  + \varDelta t }}^{-1}  \bm{H}_{t + \varDelta t} \varDelta t )^{-1}  \notag \\[10pt]
				 &\qquad \times \bm{\nabla} V_{t + \varDelta t } ( \bm{\hat{x}}_{t  + \varDelta t | t } ) \varDelta t  \label{eq:06-00-01} 
\end{align}
where $( \hat{\varSigma}_{t  + \varDelta t | t}^{-1} + \bm{H}_{t + \varDelta t}^{T} \varSigma_{\nu_{t  + \varDelta t }}^{-1}  \bm{H}_{t + \varDelta t} \varDelta t )^{-1} $ is positive-definite. 
Eq. (\ref{eq:06-00-01}) indicates that the potential term effectively gives rise to forces like classical mechanics 
and constrains $ \bm{\hat{x}}_{t  + \varDelta t | t } $ towards a local minimum or a decreasing direction of the potential function, 
if we identify the potential term with a penalty or barrier function \cite{Fletcher:2000,Nocedal_Wright:2006}. 
This fact allows us to expect that SQ using the weighted distribution can be applied to engineering problems like filtering or control, taking the appropriate potential function.

In this section, we will confirm such the expectations by the analytical or numerical examples. 
EKF is analytically derived by a quadratic potential function as the potential function representing the filtering process. 
The potential function involves the observations as an external sources, however we explain that the stochastic interpretation of the observations can be recovered like typical EKF. 
The numerical simulations are considered as the examples of control, considering the potential function as the cost function. 
It will be shown that the stochastic dynamics can be constrained towards a target direction by minimizing the potential function.
Through this section, we take 
\begin{align}
	\varDelta t = 1.
\end{align}

\subsection{Extended Kalman Filter}

To construct EKF, we consider the following quadratic potential as a penalty function: 
\begin{align}
	V_{t + 1} &= \frac{1}{2} \bm{l}_{t+1} (\bm{x}_{t + 1})^{T}  \varSigma_{\nu_{t + 1}}^{-1} \bm{l}_{t + 1} (\bm{x}_{t + 1}),  \label{eq:06-01-01} \\[10pt]
	\bm{l}_{t + 1} &= \bm{y}_{t + 1} - \bm{h}_{t+1}(\bm{x}_{t+1}). \label{eq:06-01-02} 
\end{align} 
In Eq. (\ref{eq:06-01-02}), $\bm{y}_{t + 1}$ is an external source, $\bm{h}_{t+1}(\bm{x}_{t+1})$ is a $m$-dimensional vector 
and $\varSigma_{\nu_{t + 1}}$ is an $m$-dimensional positive-definite matrix. 
Then, for the Ito process (\ref{stnc_02}), the expected dynamics is represented by the prediction process (\ref{eq:03-02})-(\ref{eq:03-03}) 
and  the update process (\ref{eq:03-04})-(\ref{eq:03-05}) which becomes as follows: 
\begin{align}
	\bm{\hat{x}}_{t +1 | t +1 } 
		&= \bm{\hat{x}}_{t +1 | t } \notag \\[5pt]
			&\quad + \hat{\varSigma}_{t +1 | t } \bm{H}_{t + 1}^{T}
				 (\varSigma_{\nu_{t + 1}}  + \bm{H}_{t + 1} \hat{\varSigma}_{t +1 | t }   \bm{H}_{t + 1}^{T} )^{-1} \notag \\[5pt]
				 	&\quad \times (\bm{y}_{t + 1} - \bm{h}_{t + 1} (\bm{\hat{x}}_{t + 1 | t})), \label{eq:06-03-01} \\[5pt]
	\hat{\varSigma}_{t +1| t +1 }
		&= \hat{\varSigma}_{t +1 | t}  \notag \\[5pt]
			&\quad - \hat{\varSigma}_{t + 1 | t }  \bm{H}_{t + 1}^{T} (\varSigma_{\nu_{t + 1}}  + \bm{H}_{t + 1} 
				 \hat{\varSigma}_{t + 1 | t }  \bm{H}_{t + 1}^{T} )^{-1} \notag \\[5pt]
				 &\quad \times \bm{H}_{t + 1} \hat{\varSigma}_{t +1 | t}. \label{eq:06-04-02}
\end{align}
It is found that (\ref{eq:03-02})-(\ref{eq:03-03}) and  (\ref{eq:06-03-01})-(\ref{eq:06-04-02}) are the formula of EKF. 
However, Eq. (\ref{eq:06-01-02}) treats $\bm{y}_{t + 1}$ as an external source and does not assume the probability distribution. 
$\varSigma_{\nu_{t + 1}} > 0$ is a constant matrix which is not directly related with $\bm{y}_{t + 1}$ as the covariance matrix.  
This is quite different point from typical EKF  \cite{Simon:2006,Saerkkae:2013} in which  $\bm{y}_{t}$ is treated as an observation given by $\bm{y}_{t+1} = \bm{h}_{t+1}(\bm{x}_{t+1}) + \bm{\xi}_{t+1}$ 
for $\bm{\xi}_{t+1} \sim \mathcal{N}(\bm{0}, \varSigma_{\nu_{t+1}})$. 
In our case, EKF can be interpreted as a process 
in which the potential constrains the expected motion of the stochastic process towards a direction satisfying 
$\bm{y}_{t + 1} \simeq \bm{h}_{t+1}(\bm{\hat{x}}_{t+1 | t})$ in which the potential function becomes the minimum.

We can also recover the stochastic interpretation  of $\bm{y}_{t + 1}$ as a special case. 
For example, by taking an approximation like 
\begin{align}
	\bm{l}_{t +1} (\bm{x}_{t + 1}) \simeq \bm{l}_{t + 1} - \bm{H}_{t + 1} (\bm{x}_{t + 1} -  \bm{\hat{x}}_{t +1 | t})
\end{align}
for Eq. (\ref{eq:06-01-02}), the weight function can be reduced to 
\begin{align}
	&e^{- U_{t + 1} (x_{t + 1} ) } = \frac{ \Omega(\bm{y}_{t +1} | \bm{x}_{t+1}) }{ \Omega( \bm{y}_{t+1} | \bm{Y}_{t })  }, \label{eq:06-05} \\[10pt]
	&\Omega(\bm{y}_{t+1} | \bm{x}_{t+1}) 
	=   \frac{ \alpha_{t+1} }{(2 \pi )^{m/2} | \varSigma_{\nu_{t+1}}  |^{1/2} } \notag \\[10pt]
		&\qquad \times e^{ - \frac{1}{2} ( \bm{y}_{t+1} - \bm{h}_{t+1}(\bm{x}_{t+1}) )^{T} \varSigma_{\nu_{t+1}}^{-1}   ( \bm{y}_{t+1} - \bm{h}_{t+1}(\bm{x}_{t+1}) ) }, \label{eq:06-06} \\[10pt]
	&\Omega( \bm{y}_{t+1} | \bm{Y}_{t }) =    \frac{ \alpha_{t+1} }{(2 \pi )^{m/2}} 
							| \varSigma_{\nu_{t+1}}  +  \bm{H}_{t+1} \hat{\varSigma}_{t+ 1| t }   \bm{H}_{t+1}^{T} |^{-1/2}  \notag \\[10pt]
							&\qquad \times e^{ - \frac{1}{2} ( \bm{l}_{t+1} )^{T} 
								( \varSigma_{\nu_{t+1}}  +  \bm{H}_{t+1} \hat{\varSigma}_{t +1| t }   \bm{H}_{t+1}^{T} )^{-1}
									 \bm{l}_{t+1} }, \label{eq:06-07}
\end{align}
where $\alpha_{t+1}$ is an arbitrary function. 
If we take $\alpha_{t+1} = 1$, 
\begin{align}
	\Omega(\bm{y}_{t+1} | \bm{x}_{t+1}) &= P(\bm{y}_{t+1} | \bm{x}_{t+1}), \\[10pt]
	\Omega( \bm{y}_{t+1} | \bm{Y}_{t }) &= P( \bm{y}_{t+1} | \bm{Y}_{t })
\end{align} 
and then we can interpret Eq. (\ref{eq:06-06}) and Eq. (\ref{eq:06-07}) as the probability densities. 
Then, $\bm{y}_{t+1}$ obeys the Gaussian distribution. This is the standard interpretation on $\bm{y}_{t+1}$ in EKF. 
In general, $\alpha_{t+1}$ is not necessarily equal to one and then Eq. (\ref{eq:06-06}) and Eq. (\ref{eq:06-07}) do not have an interpretation as a probability density. 
Our model is able to relax the assumption of the probability distribution for the observation.

\subsection{Constrained dynamics as stochastic nonlinear control}

For the applications of SQ to a nonlinear stochastic control, we will rewrite the evolution equation (\ref{eq:03-04}) to the following form: 
\begin{align}
	\bm{\hat{x}}_{t +1 | t +1 } 
		&= \bm{\hat{x}}_{t  | t } + \bm{f}_{t}(\bm{\hat{x}}_{t | t}) + B_{t} \bm{u}_{t} \label{cm-01}
\end{align}
where $\bm{u}_{t}$ is identified with the control input derived from the potential and has been defined by 
\begin{align}
	\bm{u}_{t}
		&= R_{t}^{-1}B_{t}^{T}(B_{t} R_{t}^{-1} B_{t}^{T})^{-1} \notag \\[10pt]
			&\qquad \times \hat{\varSigma}_{t + 1| t }  \bm{H}_{t + 1}^{T}
				 (\varSigma_{\nu_{t + 1}} + \bm{H}_{t + 1} \hat{\varSigma}_{t + 1 | t } \bm{H}_{t + 1}^{T} )^{-1} \notag \\[10pt]
				 	&\qquad \times  \varSigma_{\nu_{t + 1}} \bm{\nabla}_{\bm{l}_{t + 1}} V_{t + 1} (\bm{\hat{x}_{t+1|t}}) .\label{eq:03-04-02}
\end{align}
$B_{t}$ and $R_{t}$ are $m \times l$ and $l \times l$ matrix. we have assumed $R_{t}$ and $B_{t} R_{t}^{-1} B_{t}^{T}$ are regular and positive. 
Note that the covariance matrix evolves according to Eq. (\ref{eq:03-05}) as the forward process. 
This is different from a typical feedback control \cite{RFStengel:1994} and then it is not trivial whether the potential function can control the dynamics represented by Eq. (\ref{cm-01}) or not.  
To check how the control is realized, we will perform the numerical simulations by specifying the Ito process and  the potential term. 
For simplicity, we take 
\begin{align}
	&m = l,  \\[10pt]
	&B_{t} = R_{t} = 1_{m\times m}.
\end{align}

\begin{figure}[t]
  \begin{center}
    \includegraphics[keepaspectratio, width=8.4cm]{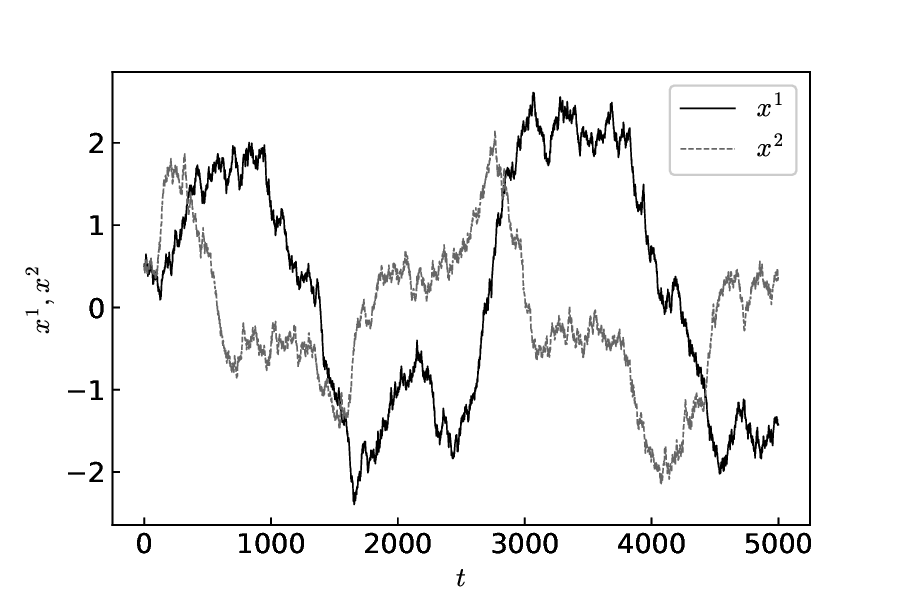}
  \end{center}
  \caption{Sample paths of the stochastic equation (\ref{stnc_02}) with the conditions (\ref{eq:inicond01})-(\ref{eq:inicond02}). }
    \label{fig:f01}
\end{figure}

In this part, we would like to consider a stochastic nonlinear control of the stochastic process (\ref{stnc_02}) with the following definitions:  
\begin{align}
	 &\bm{x}_{t} = 
	 \left(
    		\begin{array}{c}
      			x_{t}^{1} \\[10pt]
			x_{t}^{2} 
    		\end{array}
  	\right), \label{eq:07-25} \\[10pt]
	&(\bm{f}_{t}(\bm{x}_{t}))^{1} = 0.005 x_{t}^{2}, \\[10pt]
	&(\bm{f}_{t}(\bm{x}_{t}))^{2} = 0.005 \Bigl \{ (1 - (x_{t}^{1})^{2} - (x_{t}^{2})^{2} )  x_{t}^{2} \notag \\[10pt]
			 &\qquad \qquad \qquad \qquad \qquad - x_{t}^{1} + 3.0x_{t}^{2} \sin( 0.005 t ) \Bigr \}, \label{eq:07-27} \\[10pt]
	&(\bm{F}_{t})_{~1}^{1} =1, \\[10pt]
	&(\bm{F}_{t})_{~2}^{1} = 0.005, \\[10pt]
	&(\bm{F}_{t})_{~1}^{2} = 0.005 (-2 x_{t}^{1} x_{t}^{2} -1 ), \\[10pt]
	&(\bm{F}_{t})_{~2}^{2} = 1+  0.005 \Bigl \{ (-2 (x_{t}^{2})^{2}  \notag  \\[10pt]
			&\qquad \qquad + ( 1 - (x_{t}^{1})^{2} - (x_{t}^{2})^{2} ) + 3  \sin(  0.005 t  )) \Bigr \} \label{eq:07-28}
\end{align}
where $\bm{F}_{t}$ is given by Eq. (\ref{eq:03-03}) and $a$ and $b$ of $(\bm{F}_{t})^{a}_{~b}$ indicate row dimensions and column dimensions, respectively.
Fig. \ref{fig:f01} indicates sample paths of the stochastic process (\ref{stnc_02}). 
In below discussion, we will consider two cases,  ${\varSigma'}_{\nu_{t + 1}}^{-1} = 0$ and ${\varSigma'}_{\nu_{t + 1}}^{-1} \neq 0$. 

\subsubsection{${\varSigma'}_{\nu_{t + 1}}^{-1} = 0$}

In this case, the counter term vanishes, $\delta_{c} V_{t+1} = 0$ and the equation (\ref{eq:05-15}) 
implies that we are able to take $\bm{l}_{t + 1} (\bm{x}_{t + 1}) $ as a linear function on $\bm{x}_{t + 1  } $.

\begin{figure}[t]
\begin{tabular}{c}
    \includegraphics[keepaspectratio, width=8.4cm]{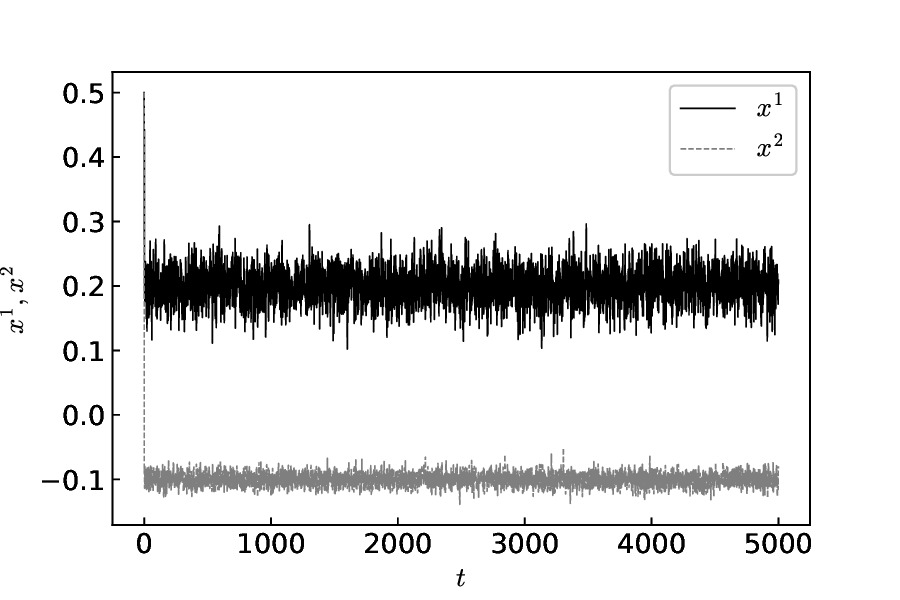}
\end{tabular}
\caption{Sample paths by sampling from the probability distribution (\ref{eq:05-27-01}) 
with the potential (\ref{eq:expotential01}) and the conditions (\ref{eq:inicond01})-(\ref{eq:inicond06}).} \label{fig:f02}
\end{figure}

\begin{figure}[t]
\begin{tabular}{c}
    \includegraphics[keepaspectratio, width=8.4cm]{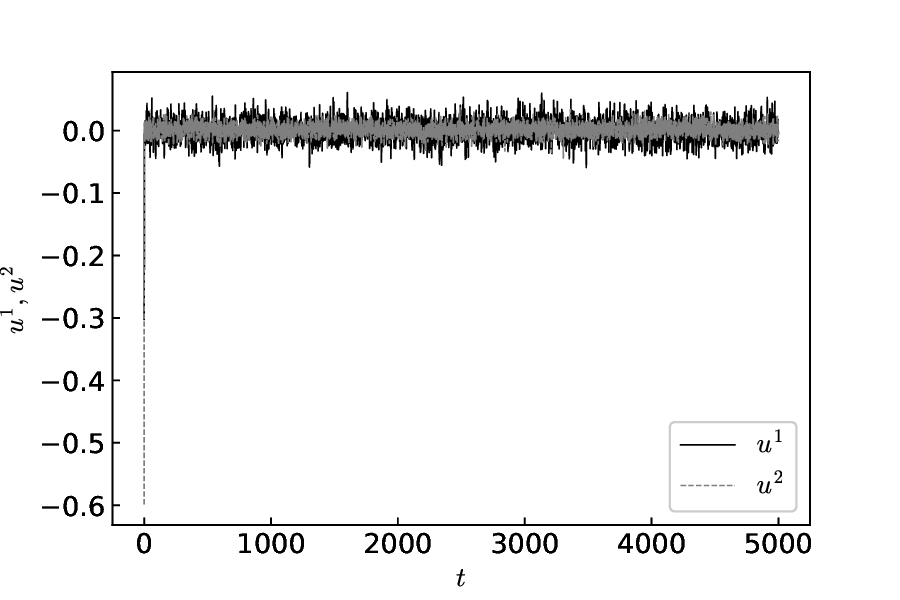}
\end{tabular}
\caption{Control inputs corresponding to Fig. \ref{fig:f02}.} \label{fig:f03}
\end{figure}

As a first example, to constrain the stochastic process towards a target $\bm{d}_{t+1}$, we will consider the following effective potential as the cost function defined by a quadratic penalty function like (\ref{eq:06-01-01}): 
\begin{align}
	V^{eff}_{t +1} (\bm{x}_{t + 1}) &= \frac{1}{2} \bm{l}_{t+1}(\bm{x}_{t + 1})^{T} \varSigma_{\nu_{t+1}}^{-1} \bm{l}_{t+1}(\bm{x}_{t + 1}), \label{eq:expotential01} \\[10pt]
	\bm{l}_{t+1} (\bm{x}_{t + 1})&= \bm{x}_{t+1} - \bm{d}_{t+1}. \label{eq:expotential02}
\end{align}
This effective potential is a positive function and has a local minimum at $\bm{d}_{t+1}$. 
By the previous discussions, the principle of least action implies $\bm{\hat{x}}_{t+1 | t+1} \rightarrow \bm{d}_{t +1}$, if we appropriately tune the parameter $\bm{g}_{t}$ and $\varSigma_{\nu_{t+1}}$. 
To numerically check this expectation, we will take the initial values as follows: 
\begin{align}
	 \bm{\hat{x}}_{0|-1} &= 
	 \left(
    		\begin{array}{c}
      			0.5 \\
			0.5 
    		\end{array}
  	\right),  \label{eq:inicond01} \\[10pt]
	\bm{g}_{t}^{-1} &=
		\begin{pmatrix}
			0.001 & 0  \\
			0 & 0.001 
		\end{pmatrix},  \label{eq:inicond02} \\[10pt]
	\hat{\varSigma}_{0|-1} &=
		\begin{pmatrix}
			1 & 0  \\
			0 & 1 
		\end{pmatrix},  \label{eq:inicond03} \\[10pt]
	\varSigma_{ \nu_{t+1}} &=
		\begin{pmatrix}
			0.001 & 0  \\
			0 & 0.0001  
		\end{pmatrix},   \label{eq:inicond04} \\[10pt]
	 \bm{d}_{t+1} &= 
	 \left(
    		\begin{array}{c}
      			0.2 \\
			-0.1 
    		\end{array}
  	\right),  \label{eq:inicond05} \\[10pt]
	\bm{H}_{t+1} &= - \bm{1}_{2 \times 2}.  \label{eq:inicond06}
\end{align}
Fig. \ref{fig:f02} and Fig. \ref{fig:f03} show sample paths of $\bm{x_{t}}$ and $\bm{u_{t}}$. 
These sample paths are given by the following steps. 
The initial control $\bm{u}_{0}$ is decided by substituting the conditions (\ref{eq:inicond01})-(\ref{eq:inicond06}) for Eq. (\ref{eq:03-04-02}).  
$\bm{x}_{1}$ is sampled by the probability distribution (\ref{eq:05-27-01}) for the conditions. 
By repeating similar steps, we can obtain the sample paths. 
Fig. \ref{fig:f02} and Fig. \ref{fig:f03} indicate that the stochastic dynamics is controlled to the local minimum of $V^{eff}_{t+1}$.

\begin{figure}[t]
\begin{tabular}{c}
    \includegraphics[keepaspectratio, width=8.4cm]{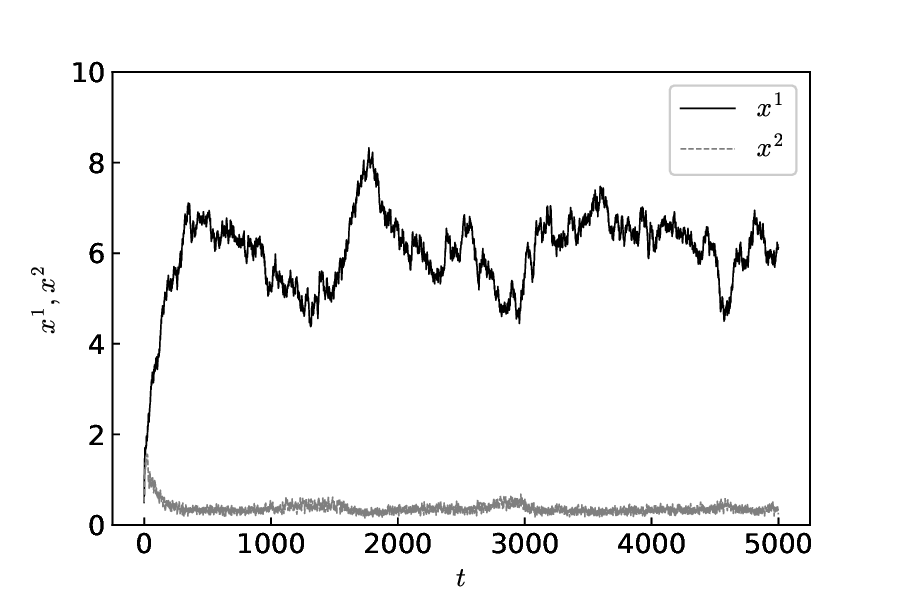}
\end{tabular}
\caption{Sample paths by sampling from the probability distribution (\ref{eq:05-27-01}). The potential is given by Eq. (\ref{eq:expotential03}) with Eq. (\ref{eq:acond}).
the sampling has used the conditions (\ref{eq:inicond01})-(\ref{eq:inicond06}) in which $\varSigma_{ \nu_{t+1}}$ is replaced by Eq. (\ref{eq:inicond07}).} \label{fig:f04}
\end{figure}

\begin{figure}[t]
\begin{tabular}{c}
    \includegraphics[keepaspectratio, width=8.4cm]{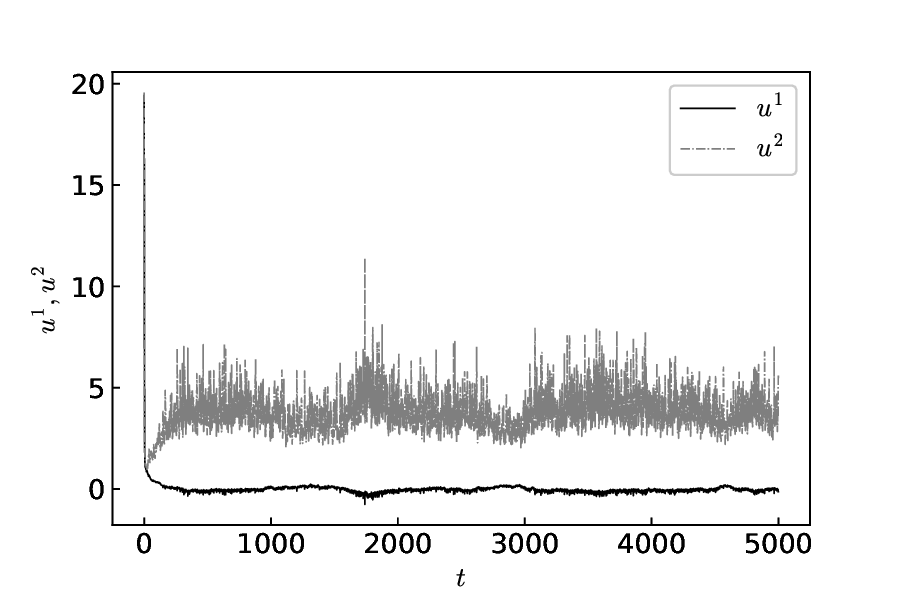}
\end{tabular}
	\caption{Control inputs corresponding to Fig. \ref{fig:f04}.} \label{fig:f05}
\end{figure}

The second example is a constrained dynamics in $\bm{x}_{t + 1} > \bm{0}$ and then the cost function is taken 
 by a potential like barrier function:  
\begin{align}
	V^{eff}_{t +1} (\bm{x}_{t + 1}) &=  -\bm{a}^{T} \ln \bm{l}_{t+1}(\bm{x}_{t + 1}) \label{eq:expotential03}
\end{align}
where $\bm{a}$ is a positive vector and 
\begin{align}
	 \ln \bm{l}_{t+1}(\bm{x}_{t + 1}) 
	 	&= \begin{pmatrix} \ln l^{1}_{t+1}(\bm{x}_{t + 1}) & \ln l^{2}_{t+1}(\bm{x}_{t + 1}) \end{pmatrix}^{T} \notag \\[10pt]
	 	&= \begin{pmatrix} \ln x^{1}_{t+1} & \ln x^{2}_{t+1} \end{pmatrix}^{T}.
\end{align}
This potential monotonically increases towards $+\infty$ for $\bm{x}_{t + 1} \rightarrow \bm{0}$ and decreases to $-\infty$ for $\bm{x}_{t + 1} \rightarrow + \infty$ . 
It is expected that the potential constrains the stochastic processes to $\bm{x}_{t + 1} > \bm{0}$. 
Using Eq. (\ref{eq:05-14}), $\varSigma_{ \nu_{t+1}}$ is  given by 
\begin{align}
	\varSigma_{ \nu_{t+1}} &=
		\begin{pmatrix}
			(\hat{x}^{1}_{t+1 | t})^{2} / a^{1} & 0  \\
			0 & (\hat{x}^{2}_{t+1 | t})^{2} / a^{2}
		\end{pmatrix}. \label{eq:inicond07}
\end{align}
$\bm{H}_{t+1}$ is given by $\bm{H}_{t+1} = - \bm{1}_{2 \times 2}$. 
The numerical simulation gives Fig. \ref{fig:f04} and Fig. \ref{fig:f05} by the same condition and manner with the first example, applying Eq. (\ref{eq:inicond07}) and 
\begin{align}
	\bm{a} = \begin{pmatrix} 10 & 10 \end{pmatrix}^{T}. \label{eq:acond}
\end{align} 
Fig. \ref{fig:f04} and Fig. \ref{fig:f05} show that the stochastic dynamics is constrained to $\bm{x}_{t + 1} > \bm{0}$,  
by the potential as barrier function.

\subsubsection{${\varSigma'}_{\nu_{t + 1}}^{-1} \neq 0$}

\begin{figure}[t]
\begin{tabular}{c}
    \includegraphics[keepaspectratio, width=8.4cm]{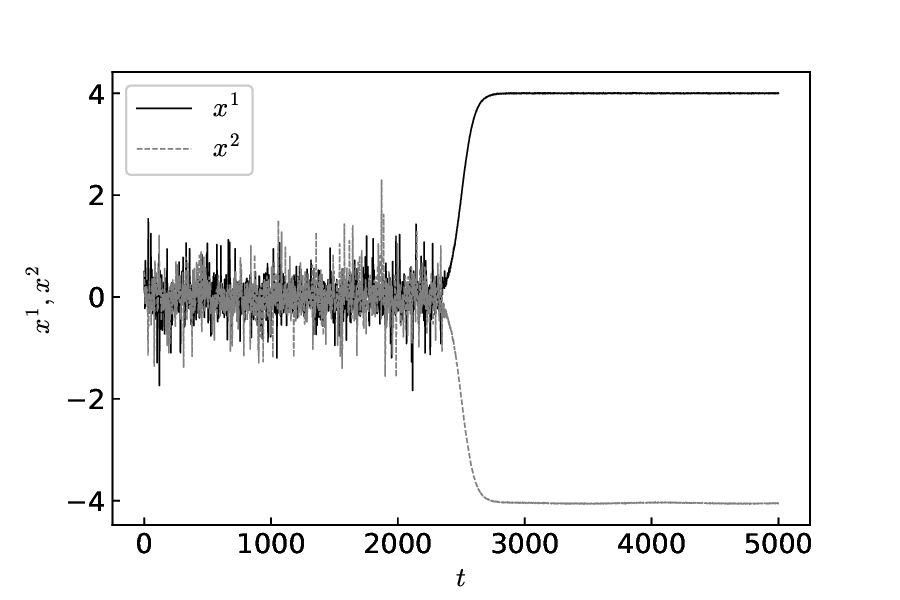}
\end{tabular}
\caption{Sample paths by sampling from the probability distribution (\ref{eq:05-27-01}). The potential is given by  Eq. (\ref{eq:expotential01}) with Eq. (\ref{eq:doublewell}).
the sampling has used the conditions (\ref{eq:inicond01})-(\ref{eq:inicond06}) 
in which $\bm{g}_{t}^{-1}$,  $\varSigma_{ \nu_{t+1}}$ and $\bm{H}_{t+1} $ are replaced by  (\ref{eq:dwcond01}), (\ref{eq:dwcond02}) and (\ref{eq:dwcond03}), respectively.} \label{fig:f06}
\end{figure}

\begin{figure}[t]
\begin{tabular}{c}
    \includegraphics[keepaspectratio, width=8.4cm]{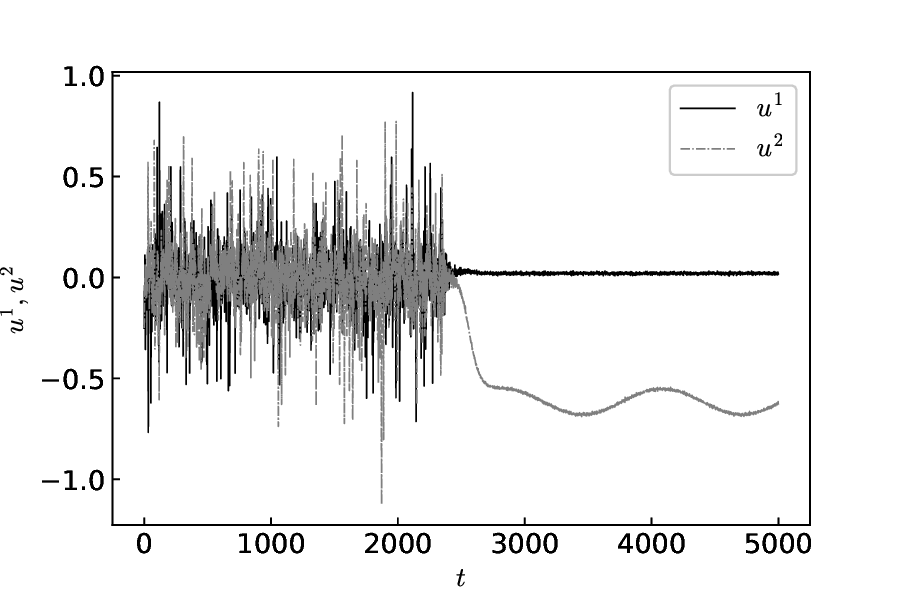}
\end{tabular}
	\caption{Control inputs corresponding to Fig. \ref{fig:f06}.} \label{fig:f07}
\end{figure}

We will consider a case in which the cost function is given by a double well potential defined by Eq. (\ref{eq:expotential01}) with 
\begin{align}
	l^{m}_{t+1} (\bm{x}_{t + 1}) &= (x^{m}_{t+1} - d^{m}_{t+1}) (x^{m}_{t+1} + d^{m}_{t+1}) \label{eq:doublewell}
\end{align}
where
\begin{align}
	 \bm{d}_{t+1} &= 
	 \left(
    		\begin{array}{c}
      			0.2 [ 1 + \tanh \{ 0.01 ( t - 2500 ) \} ]  \\[10pt]
			0.2 [ 1 + \tanh \{ 0.01 ( t - 2500 ) \} ] 
    		\end{array}
  	\right).  \label{eq:inicond03-double}
\end{align}
The double well potential has the minima at $d^{m}_{t+1}$ or $- d^{m}_{t+1}$ by Eq. (\ref{eq:doublewell}). 
The target $\bm{d}_{t+1}$ changes from $ \begin{pmatrix} 0 & 0 \end{pmatrix}^{T}$ to $ \begin{pmatrix} 0.4 & 0.4 \end{pmatrix}^{T}$ 
and then the minima of the double well potential dynamically deform, according to $\bm{d}_{t+1}$. 
Substituting Eq. (\ref{eq:expotential01}) and Eq. (\ref{eq:doublewell}) for Eq. (\ref{eq:05-15}), ${\varSigma'}_{\nu_{t + 1}}^{-1} $ is given by 
\begin{align}
	{\varSigma'}_{\nu_{t + 1}}^{-1} = 
		\begin{pmatrix}
			2 \hat{x}^{1}_{t+1 | t} l_{t+1}^{1}(\hat{x}^{1}_{t+1 | t}) & 0  \\
			0 & 2 \hat{x}^{2}_{t+1 | t} l_{t+1}^{2}( \hat{x}^{2}_{t+1 | t})
		\end{pmatrix}. 
\end{align}
The numerical simulation gives Fig. \ref{fig:f06} and Fig. \ref{fig:f07} by the same condition and manner with the previous examples, replacing the corresponding conditions by  
\begin{align}
	\bm{g}_{t}^{-1} &=
		\begin{pmatrix}
			0.5 & 0  \\
			0 & 0.5
		\end{pmatrix}, \label{eq:dwcond01} \\[10pt]
	\varSigma_{ \nu_{t+1}} &=
		\begin{pmatrix}
			0.001 & 0  \\
			0 & 0.001  
		\end{pmatrix}, \label{eq:dwcond02} \\[10pt]
	\bm{H}_{t+1} &= 
		\begin{pmatrix}
			- 2 \hat{x}^{1}_{t+1 | t} & 0  \\
			0 & - 2 \hat{x}^{2}_{t+1 | t}  
		\end{pmatrix}. \label{eq:dwcond03}
\end{align}
We find that the potential controls the stochastic processes along the dynamically changing minima. 

\section{Conclusions}\label{conclusion}

In this paper, we have proposed the discrete-time SQ using the weighted distribution as a novel and simple method to treat constrained dynamics of general Ito processes in engineering problems. 
The introduction of the weighted distribution in SQ has preserved the role of the path integral as the probability distribution 
and has naturally induced the potential term in the path integral, without imposing any assumption of the drift term. 
A second order approximation on the stochastic fluctuations to the path integral 
has given the difference equations representing the time-evolution of the expectation value and the covariance matrix for the stochastic systems. 
The difference equations have indicated that the potential term gives rise to forces constraining the stochastic dynamics. 
The difference equations has derived EKF by the potential function as the penalty function.
Although the penalty function has involved observations as just external sources,  we have also explained a case where such the stochastic interpretation of the observations is recovered as the typical definition of EKF. 
The potential function as a barrier or penalty function has been applied to model the nonlinear stochastic control. 
It has been numerically shown that the nonlinear stochastic processes can be constrained  towards a minimum or a decreasing direction of the potential function.

For more realistic applications, how to accept constraints on control inputs induced from the potential term also remains to be considered. 
A hyperparameter optimization is also required since the stability of a given stochastic system depends on the hyperparameters. 
The computation on the covariance matrix has the high cost for a large-scale and high-dimensional system. 
Thus, how to reduce the computational cost is remaining problem to be solved in order to be more powerful tool for real world. 
It is also interesting to consider parameter estimation because EKF is able to estimate parameters of a machine learning model \cite{Singhal_Wu:1988}.

\begin{acknowledgements}
The author would like to thank Koichi Hamada, Shinichi Takayanagi and Hirofumi Yamashita for valuable discussions and comments about early ideas on this paper at a seminar. 
The author is also grateful to Yu Higashiyama and Takashi Kawasaki for the opportunity and the support of the seminar at Fringe81 Co., Ltd..
\end{acknowledgements}

\appendix

\section{Matrix inversion lemma and product rules for determinant}\label{A01}

$A$ and $D$ are invertible square matrices, and $B$ and $C$ matrices may or may not be square. 
The matrix inversion lemma  and the product rules for determinant are given as follows \cite{Simon:2006}: 
\begin{align}
	&(A + B D^{-1} C)^{-1} = A^{-1} - A^{-1} B (D + C A^{-1} B)^{-1} C A^{-1}, \\[10pt]
	&\left|
    		\begin{array}{cc}
      			A & B  \\
     			C & D
    		\end{array}
  \right| = |A - B D^{-1}C| |D| = |A| |D-C A^{-1}B|.
\end{align}

\section{Identities}\label{A02}

Applying the matrix inversion lemma in Appendix \ref{A01}, we can show the following equations: 
\begin{align}
	&e^{ - U_{t + \varDelta t} \varDelta t } p(x_{t + \varDelta t}, t + \varDelta t  )  \notag \\[10pt]
		& =  \frac{ N_{t + \varDelta t } e^{ - \mathcal{N}_{t + \varDelta t } \varDelta t }  | \hat{\varSigma}_{t + \varDelta t | t + \varDelta t } | ^{1/2}  }
				{  | \hat{\varSigma}_{t + \varDelta t | t } | ^{1/2} } \notag \\[10pt]
		        &\quad \times \frac{ 1  }{ (2 \pi )^{m/2} | \hat{\varSigma}_{t + \varDelta t | t + \varDelta t } | ^{1/2} } \notag  \\[10pt]
			&\quad \times e^ {- \frac{1}{2} (\bm{x}_{t + \varDelta t } - \bm{\hat{x}}_{t + \varDelta t | t + \varDelta t } )^{T} 
			\hat{\varSigma}_{t + \varDelta t | t + \varDelta t  }^{ -1} 
			(\bm{x}_{t + \varDelta t} - \bm{\hat{x}}_{t + \varDelta t | t + \varDelta t} )  } \notag  \\[10pt]
		&= P( x_{t + \varDelta t}, t + \varDelta t  )  \label{eq:A02-01}
\end{align}
where we have defined the following equations: 
\begin{align}
	\bm{\hat{x}}_{t + \varDelta t | t  + \varDelta t } 
		&=  \bm{\hat{x}}_{t  + \varDelta t | t } \notag \\[10pt]
			&\quad + ( \hat{\varSigma}_{t  + \varDelta t | t }^{-1} + \bm{H}_{t + \varDelta t}^{T} \varSigma_{\nu_{t + \varDelta t}}^{ -1}  \bm{H}_{t + \varDelta t} \varDelta t )^{-1}  \notag \\[10pt]
				 &\qquad \times \bm{H}_{t  + \varDelta t }^{T}
				 \bm{\nabla}_{\bm{l}_{t + \varDelta t}} V_{t + \varDelta t} \varDelta t  \label{eq:A02-02-02}  \\[10pt]
		&= \bm{\hat{x}}_{t  + \varDelta t | t } \notag \\[10pt]
			&\quad + \hat{\varSigma}_{t  + \varDelta t | t } \bm{H}_{t  + \varDelta t }^{T} \notag \\[10pt]
				 &\qquad \times (\varSigma_{\nu_{t  + \varDelta t}} / \varDelta t  + \bm{H}_{t + \varDelta t} \hat{\varSigma}_{t  + \varDelta t | t}   \bm{H}_{t + \varDelta t}^{T} )^{-1} \notag \\[10pt]
				 	 &\qquad \times \varSigma_{\nu_{t  + \varDelta t }} \bm{\nabla}_{\bm{l}_{t + \varDelta t}} V_{t + \varDelta t}, \label{eq:A02-02-04} 
\end{align}
\begin{align}
	\hat{\varSigma}_{t  + \varDelta t | t  + \varDelta t}
		&= ( \hat{\varSigma}_{t  + \varDelta t| t }^{-1} + \bm{H}_{t + \varDelta t}^{T} 
			\varSigma_{\nu_{t + \varDelta t}}^{-1}  \bm{H}_{t + \varDelta t} \varDelta t )^{-1}   \label{eq:A02-03-02}   \\[10pt] 
		&= \hat{\varSigma}_{t  + \varDelta t | t } \notag \\[10pt]
			&\quad - \hat{\varSigma}_{t  + \varDelta t| t} \bm{H}_{t + \varDelta t}^{T} \notag \\[10pt]
				&\qquad \times (\varSigma_{\nu_{t + \varDelta t}} / \varDelta t + \bm{H}_{t + \varDelta t} \hat{\varSigma}_{t + \varDelta t | t }  \bm{H}_{t + \varDelta t}^{T}  )^{-1} \notag \\[10pt]
					&\qquad \times \bm{H}_{t + \varDelta t} \hat{\varSigma}_{t  + \varDelta t| t}, \label{eq:A02-03} \\[10pt]
	| \varSigma_{\nu_{t + \varDelta t}} / \varDelta t &+  \bm{H}_{t + \varDelta t} \hat{\varSigma}_{t + \varDelta t | t } 
		 \bm{H}_{t + \varDelta t}^{T} |^{1/2} \notag \\[10pt]
		 	&\times |\varSigma_{\nu_{t + \varDelta t}}/ \varDelta t|^{ -1/2}  |\hat{\varSigma}_{t  + \varDelta t| t } |^{ -1/2} \notag \\[10pt]
	 &= | \hat{\varSigma}_{t + \varDelta t | t }^{-1} + \bm{H}_{t + \varDelta t }^{T} \varSigma_{\nu_{t  + \varDelta t }}^{-1}  \bm{H}_{t  + \varDelta t} \varDelta t  |^{1/2}.  \label{eq:A02-04}
\end{align}

%
\section*{Conflict of interest}
The authors declare that they have no conflict of interest. 
This manuscript is an extended version of author's presentation on stochastic process, 
when the author was working at Fringe81 Co., Ltd..  
The manuscript is not related to the business.



\end{document}